\documentclass[aps,prb,twocolumn,showpacs,letterpaper,superscriptaddress]{revtex4-1}
\usepackage{graphicx}
\usepackage{multirow}
\usepackage{amsmath}
\usepackage{color}
\usepackage[normalem]{ulem}

\begin{document}

\title{Optical conductivity renormalization of graphene on  SrTiO$_{3}$ due to resonant excitonic effects mediated by Ti 3\textit{d} orbitals}
\affiliation{Singapore Synchrotron Light Source, National University of Singapore, 5 Research Link, Singapore 117603}
\affiliation{Department of Physics, National University of Singapore, Singapore 117542}
\affiliation{NUSNNI-NanoCore, National University of Singapore, Singapore 117576}
\affiliation{Graphene Research Centre, Faculty of Science, National University of Singapore, Singapore 117546}
\affiliation{Department of Chemistry, National University of Singapore, Singapore 117543}
\affiliation{Department of Electrical and Computer Engineering, National University of Singapore, Singapore 117576}
\affiliation{Center for Biomolecular Nanotechnologies @UNILE, Istituto Italiano di Tecnologia, Via Barsanti, I-73010 Arnesano, Italy}
\affiliation{National Nanotechnology Laboratory (NNL), Istituto Nanoscienze-CNR,
Via per Arnesano 16, I-73100 Lecce, Italy }
\affiliation{Jurusan Fisika-FMIPA-Universitas Gadjah Mada, BLS 21 Sekip Utara, Yogyakarta, Indonesia 55281}

\author{Pranjal Kumar Gogoi}
\affiliation{Singapore Synchrotron Light Source, National University of Singapore, 5 Research Link, Singapore 117603}
\affiliation{Department of Physics, National University of Singapore, Singapore 117542}
\affiliation{NUSNNI-NanoCore, National University of Singapore, Singapore 117576}
\author{Paolo E. Trevisanutto}
\affiliation{Singapore Synchrotron Light Source, National University of Singapore, 5 Research Link, Singapore 117603}
\affiliation{Department of Physics, National University of Singapore, Singapore 117542}
\affiliation{Graphene Research Centre, Faculty of Science, National University of Singapore, Singapore 117546}
\affiliation{Center for Biomolecular Nanotechnologies @UNILE, Istituto Italiano di Tecnologia, Via Barsanti, I-73010 Arnesano, Italy}
\author{Ming Yang}
\affiliation{Singapore Synchrotron Light Source, National University of Singapore, 5 Research Link, Singapore 117603}
\affiliation{Department of Physics, National University of Singapore, Singapore 117542}
\author{Iman Santoso}
\affiliation{Singapore Synchrotron Light Source, National University of Singapore, 5 Research Link, Singapore 117603}
\affiliation{Department of Physics, National University of Singapore, Singapore 117542}
\affiliation{Graphene Research Centre, Faculty of Science, National University of Singapore, Singapore 117546}
\affiliation{Jurusan Fisika-FMIPA-Universitas Gadjah Mada, BLS 21 Sekip Utara, Yogyakarta, Indonesia 55281}
\author{Teguh Citra Asmara}
\affiliation{Singapore Synchrotron Light Source, National University of Singapore, 5 Research Link, Singapore 117603}
\affiliation{Department of Physics, National University of Singapore, Singapore 117542}
\affiliation{NUSNNI-NanoCore, National University of Singapore, Singapore 117576}
\author{Aleksandrs Terentjevs}
\affiliation{Center for Biomolecular Nanotechnologies @UNILE, Istituto Italiano di Tecnologia, Via Barsanti, I-73010
Arnesano, Italy}
\author{Fabio Della Sala}
\affiliation{Center for Biomolecular Nanotechnologies @UNILE, Istituto Italiano di Tecnologia, Via Barsanti, I-73010 Arnesano, Italy}
\affiliation{National Nanotechnology Laboratory (NNL), Istituto Nanoscienze-CNR,
Via per Arnesano 16, I-73100 Lecce, Italy }

\author{Mark B. H. Breese}
\affiliation{Singapore Synchrotron Light Source, National University of Singapore, 5 Research Link, Singapore 117603}
\affiliation{Department of Physics, National University of Singapore, Singapore 117542}

\author{T. Venkatesan}
\affiliation{Department of Physics, National University of Singapore, Singapore 117542}
\affiliation{NUSNNI-NanoCore, National University of Singapore, Singapore 117576}
\affiliation{Department of Electrical and Computer Engineering, National University of Singapore, Singapore 117576}

\author{Yuan Ping Feng}
\affiliation{Department of Physics, National University of Singapore, Singapore 117542}

\author{Kian Ping Loh}
\affiliation{Graphene Research Centre, Faculty of Science, National University of Singapore, Singapore 117546}
\affiliation{Department of Chemistry, National University of Singapore, Singapore 117543}

\author{Antonio H. Castro Neto}
\affiliation{Department of Physics, National University of Singapore, Singapore 117542}
\affiliation{Graphene Research Centre, Faculty of Science, National University of Singapore, Singapore 117546}
\author{Andrivo Rusydi}
\email{phyandri@nus.edu.sg}
\affiliation{Singapore Synchrotron Light Source, National University of Singapore, 5 Research Link, Singapore 117603}
\affiliation{Department of Physics, National University of Singapore, Singapore 117542}
\affiliation{NUSNNI-NanoCore, National University of Singapore, Singapore 117576}
\date{\today}
%
%
\begin{abstract}

We present evidence of a drastic renormalization of the optical conductivity of graphene on SrTiO$_3$  resulting in almost full transparency in the ultraviolet region.  These findings are attributed to resonant excitonic effects further supported by \emph{ab initio} Bethe-Salpeter equation and density functional theory calculations. The ($\pi$,$\pi$*)-orbitals of graphene and Ti-3\textit{d} $t_{2g}$ orbitals of SrTiO$_3$ are strongly hybridized and the interactions of electron-hole states residing in those orbitals play dominant role in the graphene optical conductivity. These interactions are present much below the optical band gap of bulk SrTiO$_3$. These results open a possibility of manipulating interaction strengths in graphene via \textit {d}-orbitals which could be crucial for optical applications.

\end{abstract}


\maketitle
\section{Introduction}
Graphene, the thinnest material to be successfully isolated, manifests prominent many-body effects of electron-electron (e-e) and electron-hole (e-h) interactions which can be manipulated by using substrate materials \cite{CastroNeto, Jang, Peres, Kotov}. The role of interacting quasi-particles in graphene, particularly in the form of considerable e-e and e-h interactions has been revealed by recent reports such as renormalization of the Fermi velocity with distortion of the Dirac cone \cite {Elias}, fractional quantum Hall effect \cite{Bolotin, Du} and prominent excitonic effects and interactions, which occur even at high energy in optical absorbance spectra \cite{Nair, Yang, Trevisanutto, Wehling, Kravets, Mak11, Chae, Gogoi, Santoso11, Santoso14, Mak14}. In the context of many-body effects, optical conductivity measurements of graphene allow one to reveal the roles of both e-e and particularly e-h interactions as optical transitions involve the creation of concomitant e-h states.

For such studies graphene has been typically measured either on wide band gap substrates \cite{Mak11, Kravets, Gogoi, Santoso11} or in the free-standing configuration \cite{Chae, Nair}. However graphene on a low (\textless$\sim$1 eV) or intermediate band gap substrate (up to $\sim$5 eV) could be an intriguing system to study as the interfacial optical processes both in graphene and the substrate (which is absent for a wide band gap substrate) may affect the intrinsic characteristics of each other. An ideal example of an intermediate band gap substrate is SrTiO$_3$ (with band gap of  $\sim$3.2 eV). In particular,  SrTiO$_3$ is regarded as a model 3$d^0$ system and has been widely used as a substrate generating an interface exhibiting new fundamental phenomena \cite {Asmara, Hwang} in recent times. An optical conductivity study of Graphene-SrTiO$_3$ interface allows us to reveal how orbitals in graphene hybridize and interact with such Ti-3$d$ orbitals of transition metal oxides.

In this work we show new phenomenon of anomalous renormalization of graphene optical  conductivity using  SrTiO$_3$ as the substrate. The optical conductivity of the graphene layer shows novel features unlike the intrinsic case particularly with almost full quenching beyond $\sim$3.2 eV. These observations are supported by \emph{ab initio} many body perturbation theory  Bethe-Salpeter Equation (BSE) calculations which incorporates e-h effects.  Our results provide evidence of strong interactions between the  Ti-\textit{3d}  and graphene \textit{2p$_z$} orbitals.

\section{Experimental Technique and Samples}

Reflective spectroscopic ellipsometry (SE) is used to measure the optical conductivity of graphene. Reflective SE is capable of probing physical properties of extremely thin materials like graphene or other ultrathin flms on a bulk substrate \cite {Gogoi, Santoso14, Asmara}. Spectroscopic ellipsometry measurements are performed using a SENTECH SE 850 ellipsometer as reported in previous studies \cite{Gogoi, Santoso14}.  The reliable data range is limited to 0.5 - 5.2 eV due to the limitation of the micro-focus probes. Here in this work, data from 0.5 - 5.2 eV are reported for all measurements. The substrates used in this work: SrTiO$_3$ (100) and SiO$_2$/Si are homogeneous, isotropic and atomically flat. Spectroscopic ellipsometry measurements of $(\Psi, \Delta)$ at different spots on the substrates are found to be identical for individual incident angles. For modelling as well as extraction of optical parameters, $(\Psi, \Delta)$ measured at one spot for different incident angles are used. Use of simultaneous fitting of several incident angle data is crucial for the uniqueness of the final results. 

\begin{figure}
\includegraphics[width=\columnwidth]{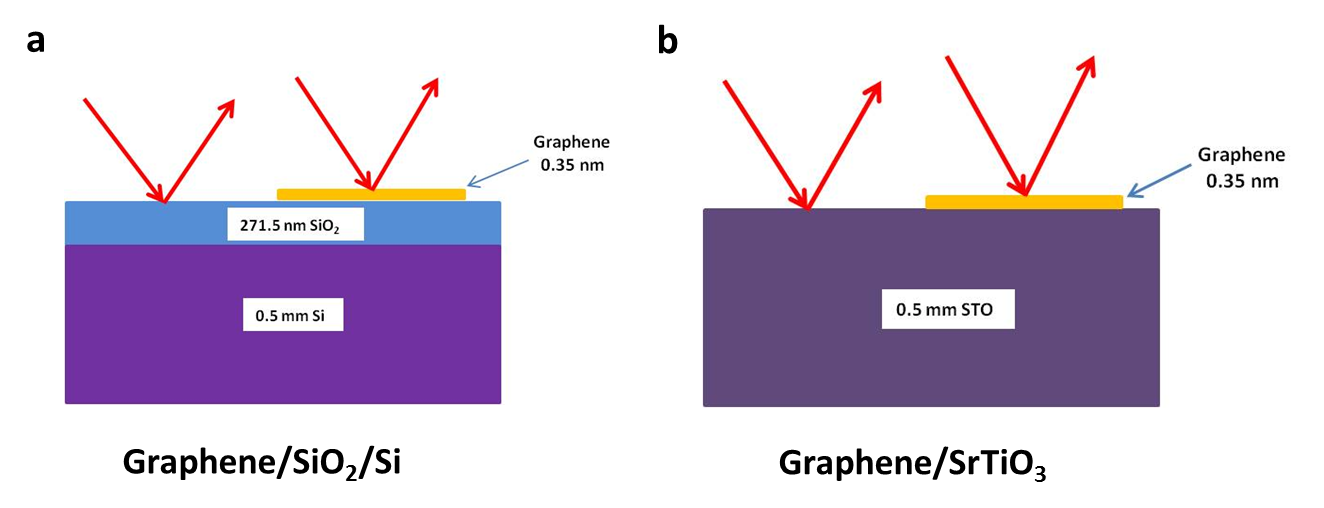}
\caption{\label{fig1} (Color online).
Schematic cross section of the graphene samples under study (a) graphene/SiO$_2$/Si and (b) graphene/SrTiO$_3$. (The thicknesses are not to scale.)}
\end{figure}

The graphene samples (procured from Graphene Square Inc.), which are prepared by chemical vapour deposition, are transferred to substrates using wet transfer method \cite{Bae}. The graphene layers have been carefully positioned during transfer to the substrates such that they cover approximately one half of the top side of the substrate as shown in Fig. \ref{fig1}. This is to make sure that the possible differences in optical constants of the substrate which could arise due to measurements on separate substrates are eliminated and exactly the same substrate optical constants could be used in the analysis of graphene/substrate in the individual cases.

\begin{figure}
\includegraphics[width=2.3 in]{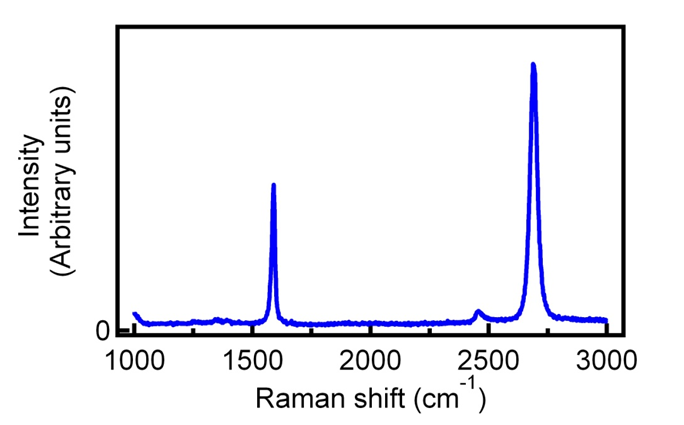}
\caption{\label{fig2} (Color online).
Raman Spectra of graphene/SiO$_2$/Si  with 514.5 nm laser.}
\end{figure}

\begin{figure*}
\includegraphics[width= 4.5in]{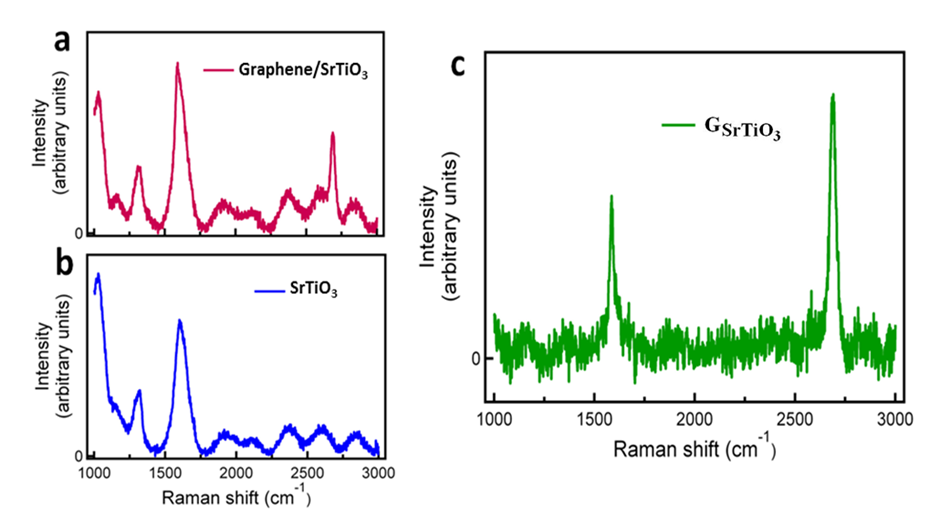}
\caption{\label{fig3} (Color online).
(a) Raman spectra of graphene/SrTiO$_3$ with 514.5 nm laser. (b) Raman spectra of  SrTiO$_3$ with 514.5 nm laser. (c)  (Normalized) Raman Spectra of G$_{SrTiO_3}$ with 514.5 nm laser.}
\end{figure*}

Raman measurement using a 514.5 nm laser performed on the graphene/SiO$_2$/Si sample shows distinct single layer characteristics \cite{Ferrari} as well as negligible defects in the graphene layer as plotted in Fig. \ref{fig2}. Raman measurements performed on graphene/SrTiO$_3$ and  SrTiO$_3$ substrate part only are shown in Fig. \ref{fig3}(a) and \ref{fig3}(b) respectively. Figure \ref{fig3}(c) shows the normalized Raman shift only for the graphene layer on SrTiO$_3$ (G$_{SrTiO_3}$). The shape and relative intensities of G and 2D peaks in Fig. \ref{fig3}(c) clearly show the single layer characteristics \cite{Ferrari}. The noisy baseline is due to subtraction of comparable small numbers and as a result it is hard to resolve if there is any defect peak (D peak) present. However the defect contribution is known to be negligible from Raman measurement of graphene/SiO$_2$/Si  as can be seen in Fig. \ref{fig2}. All the transferred graphene layers are taken from the same larger piece of graphene on copper foil.

Similar results are obtained for SE measurements  performed for graphene on  both SrTiO$_3$  (100) and SrTiO$_3$ (110). It is noteworthy that the bulk single crystalline SrTiO$_3$  used here is not ferroelectric. All reported measurements are performed on three different samples of graphene on SrTiO$_3$  (100) and the results are reproducible.

\section{Spectroscopic ellipsometry data analysis}

Multilayer modelling which takes into account reflections at each interface through Fresnel  coefficients \cite{Hecht} is used for simultaneous fitting of data measured at multiple incident angles. We have used Drude-Lorentz oscillators for the fittings \cite{Kuzmenko, Fujiwara, Tompkins}. The graphene layer has been assumed to be flat and isotropic as reported in similar studies \cite{Nelson}. For graphene/SiO$_2$/Si as well as for the substrate alone (SiO$_2$/Si) we have used 70$^\circ$, 60$^\circ$, 50$^\circ$, 40$^\circ$ incident angle data. For graphene/SrTiO$_3$ as well as the substrate alone (SrTiO$_3$) we have used 70$^\circ$, 65$^\circ$, 60$^\circ$ incident angle data.

\begin{figure*}
\includegraphics[width= 5.5in]{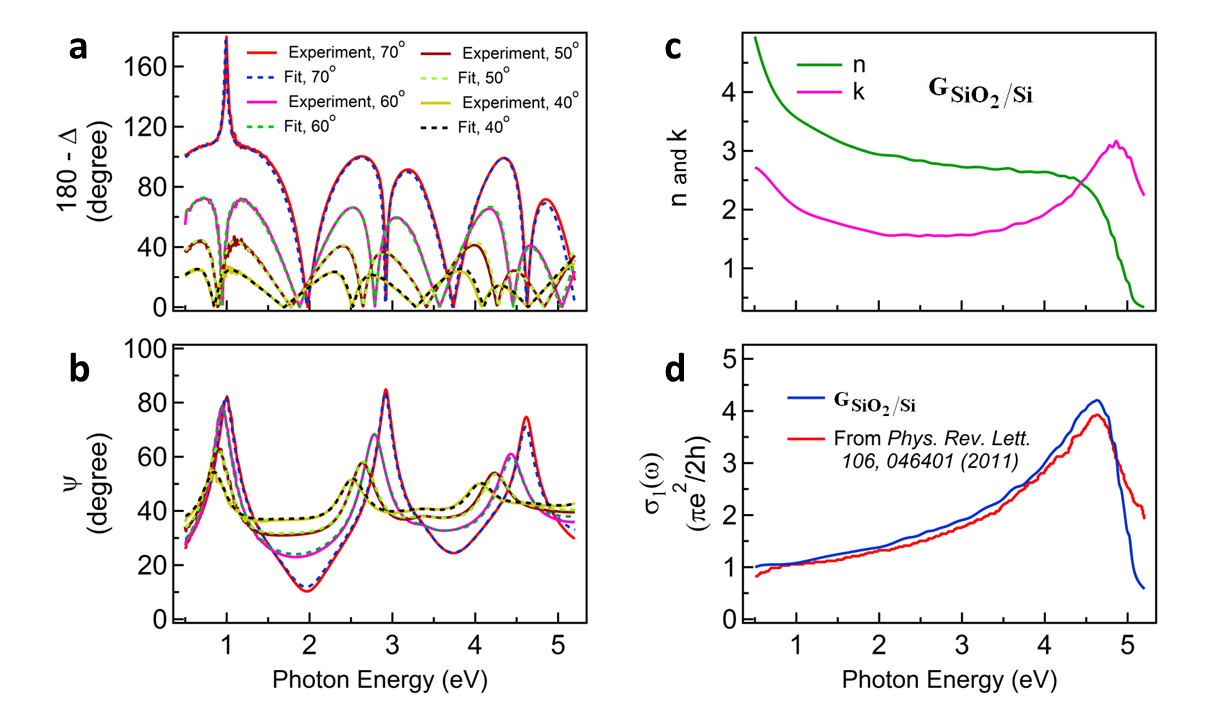}
\caption{\label{fig4} (Color online).
(a) ($180 - \Delta$) data and fit for graphene/SiO$_2$/Si. (b) $\Psi$ data and fit for graphene/SiO$_2$/Si. (c) Extracted $(n, k)$ for G$_{SiO_2/Si}$.  (d) Extracted $\sigma_1(\omega)$ for G$_{SiO_2/Si}$ plotted with reported previous result \cite{Mak11} obtained from measurements on exfoliated graphene.}
\end{figure*}

\begin{figure}
\includegraphics[width=3in]{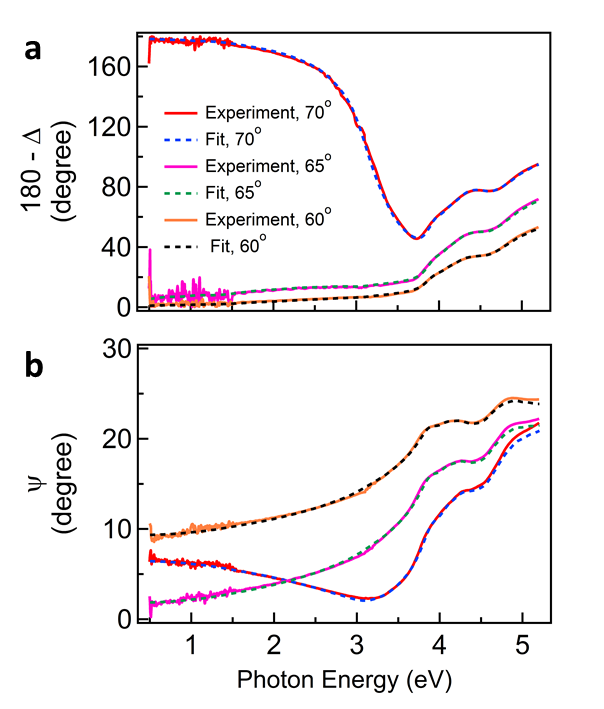}
\caption{\label{fig5} (Color online).
(a) ($180-\Delta$) data and fit for graphene/SrTiO$_3$. (b) $\Psi$ data and fit for graphene/SrTiO$_3$.}
\end{figure}

To start the analysis of graphene/SiO$_2$/Si we first fit the measured values of  $(\Psi, \Delta)$ on a Si substrate to extract the $(\varepsilon_1, \varepsilon_2)$  of Si. These $(\varepsilon_1, \varepsilon_2)$ have been used to fit the  $(\Psi, \Delta)$ for SiO$_2$/Si in turn. Now, using the extracted thickness (of the SiO$_2$ layer) as well as $(\varepsilon_1, \varepsilon_2)$ of SiO$_2$ and Si, $(\Psi, \Delta)$ measured on graphene/SiO$_2$/Si are fitted to extract the optical constants of the graphene layer on SiO$_2$/Si (G$_{SiO_2/Si}$).

Figure \ref{fig4}(a) and \ref{fig4}(b) show the data and fit of $(180 - \Delta)$  and $\Psi$ for graphene/SiO$_2$/Si measured on the graphene covered part of the substrate  as shown in Fig. \ref{fig1}(a). In our fittings, we have used one monolayer thickness of 3.35 \AA \,for graphene. It may be mentioned that for the best fit we have to use a Cauchy layer of thickness 1 nm between graphene and the substrate similar to what has been reported previously \cite{Kravets}. The extracted $(n, k)$ for the graphene layer is plotted in Fig. \ref{fig4}(c) which is similar to other reports on exfoliated as well as CVD graphene \cite{Nelson, Weber}. Figure \ref{fig4}(d) shows the extracted  $\sigma_1(\omega)$  from fitting and also the comparison with reported $\sigma_1(\omega)$ for exfoliated graphene found using reflectivity measurements \cite{Mak11}. The slight higher value of $\sigma_1(\omega)$  in our result may be attributed to the presence of some amount of bilayer areas (below 5\%) in our CVD graphene (which is normally observed for CVD graphene \cite{Bae}).

The SrTiO$_3$ substrate has not been treated for any preferential termination \cite{Koster}. So it is expected that both SrO and TiO terminations are there on the surface equally. The roughness of the surface is found to be less than $\sim$5 \AA \,by AFM measurements. This atomically flat substrate is reasonably modeled using a flat underlying substrate with graphene sitting on top in our analysis. The $(\Psi, \Delta)$ data measured on the substrate part of the sample as shown in Fig. \ref{fig1}(b) can be directly converted to the pseudo--dielectric function $(\langle\varepsilon_1\rangle, \langle\varepsilon_2\rangle)$. For an ideal isotropic flat bulk substrate the pseudo--dielectric function approaches the true dielctric function $(\varepsilon_1, \varepsilon_2)$ \cite{Tompkins, Fujiwara}. Since the SrTiO$_3$ substrate used here is isotropic and atomically flat, we have used the extracted pseudo-dielectric function as the true dielectric function for all our analysis. This  $(\varepsilon_1, \varepsilon_2)$ is used to model  and extract  the Drude-Lorentz oscillator parameters later to be used for fitting of data measured on graphene layer supported on this substrate. 

Figure \ref{fig5} shows the $(180 - \Delta)$ and $\Psi$ data and fit respectively for graphene/SrTiO$_3$. It may be mentioned that for the best fit a Cauchy layer of thickness 2 \AA  \,has to be used in this case  in between graphene layer and SrTiO$_3$ substrate. The extracted $\sigma_1(\omega)$ is plotted in Fig. \ref{fig8}(b).
The most important results of this combined experimental and theoretical study are based on the comparison of $\sigma_1(\omega)$ for G$_{SiO_2/Si}$ and G$_{SrTiO_3 }$ as shown in Fig. \ref{fig8}. These crucial aspects will be discussed in detail in Section \ref{sec6}.

\section{Ground State: Density functional theory (DFT) calculation details}\label{sec4}

All ground state calculations are first carried out by using density functional theory (DFT) based Vienna \textit{ab initio} simulation package (VASP) with the Perdew--Burke--Ernzerhof (PBE) approximation for the exchange-correlation functional \cite{Kresse9347, Kresse9348}. The frozen-core all-electron projector-augmented wave (PAW) \cite{Blochl} method, as implemented in VASP, is used. The cut-off energy for the expansion of plane-wave basis is 500 eV. The effective on-site Coulomb correction \cite{Himmetoglu} ($U=5.0$ eV and $J=0.64$ eV) is applied to $d$ orbital electrons in Ti atom in accordance to the previous works on SrTiO$_3$ \cite{Mizokawa, Okamoto}.  For SrTiO$_3$ bulk and pristine graphene, 8$\times$8$\times$8 and 12$\times$12$\times$1  $k$-point meshes are used, respectively.  The electronic convergence is set to 1.0$\times$10$^{-6}$ eV, and the force on each atom is optimized smaller than 0.01 eV/\AA\, for all calculations including the graphene/SrTiO$_3$ interface. Based on these parameters, the calculated lattice constant and band gap of SrTiO$_3$ are 3.976 \AA \,and 2.39 eV, respectively, and the lattice constant for graphene is 2.46 \AA,  in good agreement with previous studies \cite{Benthem}.

\begin{figure}
\includegraphics[width=1.8 in]{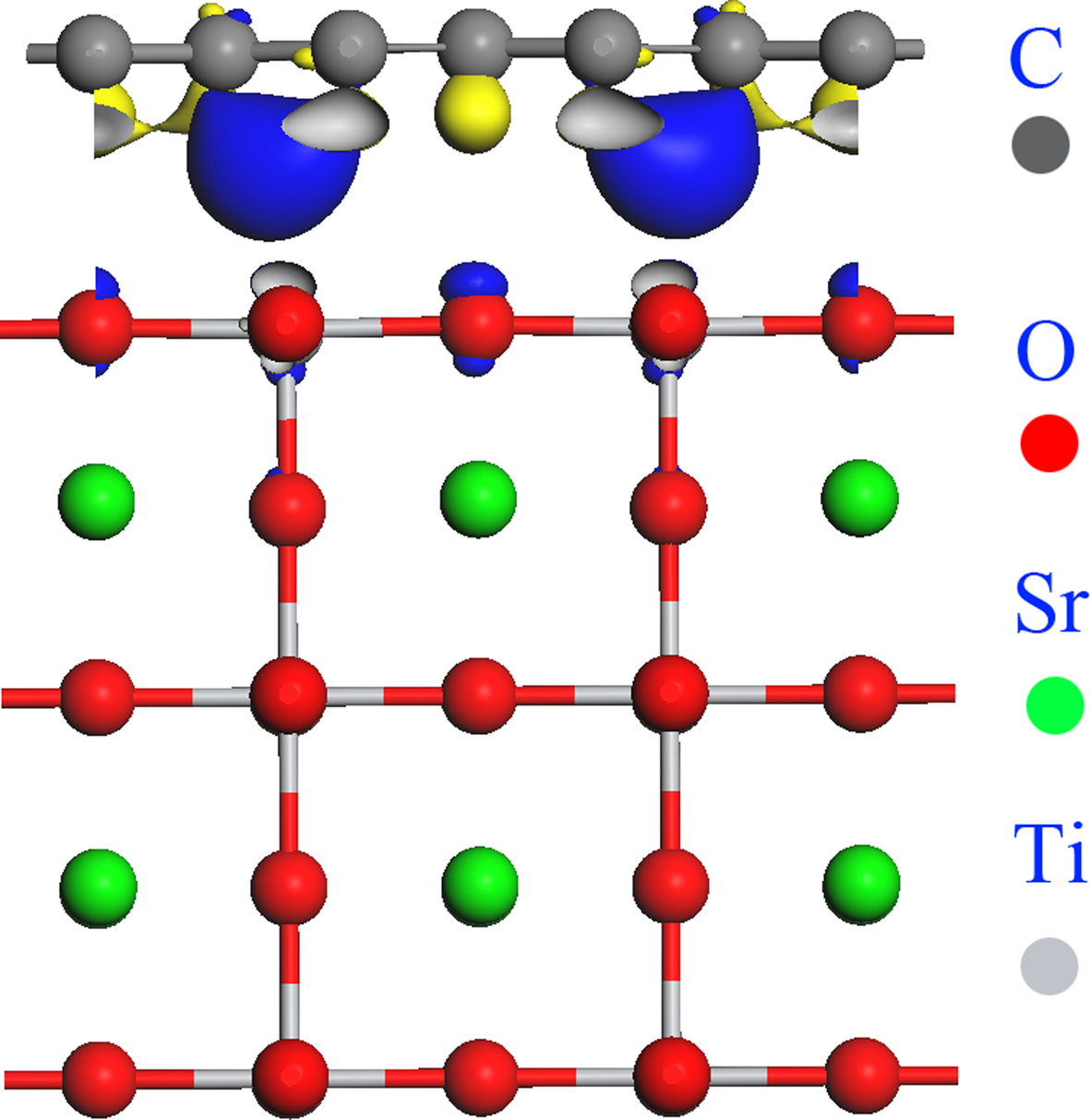}
\caption{\label{fig6} (Color online).
Interfacial charge distribution plot.}
\end{figure}

To model graphene on SrTiO$_3$ substrate, $(\sqrt{3}\times 3)$ graphene supercell is  strained and placed on $(1 \times2)$ SrTiO$_3$ (100) surface with TiO$_2$ termination and 5 layers of thickness, in which 7.14\% compressive and 7.19\% tensile strains are applied along graphene $a$ and $b$ lattice directions, respectively, to match the lattice constants of SrTiO$_3$ substrate. The vacuum region in the interface models was set to 15 \AA \,to minimize Coulomb interactions of neighbour surfaces, and 8$\times$4$\times$1 k-point meshes are used for the interface structures. The bottom two layers of SrTiO$_3$ substrate are fixed during the relaxation process, and van der Waals effect between SrTiO$_3$ and graphene are included also \cite{Grimme}. In addition, the charge redistribution at the interface is analysed by using Bader script \cite{Henkelman} as shown in Fig. \ref{fig6}.

The optimized most energetically favourable interface structure is shown in Fig. \ref{fig6}, where the shortest distance between graphene and SrTiO$_3$ surface is about 2.83 \AA. This distance is much larger than the bond length of potential interfacial covalent C-O bond ($\sim$1.42 \AA \,for single bond in CO) or ionic Ti-C bond ($\sim$2.18 \AA \,for bulk TiC). Moreover, the charge transfer at the interface is found weak. The charge density difference, $\Delta\rho$  suggests the charge transfer at the interface, which is defined as

\begin{equation}
 \Delta\rho = \rho_{graphene/SrTiO_3} - (\rho_{graphene} + \rho_{SrTiO_3})
\end{equation}

where $\rho_{graphene/SrTiO_3}$ is the charge density of graphene/SrTiO$_3$, and $\rho_{graphene}$ and $\rho_{SrTiO_3}$ are the corresponding charge densities of graphene and SrTiO$_3$ substrate, respectively. As shown in in Fig. \ref{fig6} with a small iso-surface value of $4.0 \times 10^{-4} e/{\emph{\AA}}^3$, the excess charge density can be seen near O atom at SrTiO$_3$ surface, while depleted charge density is at C atoms of graphene.  The calculation of Bader charge redistribution shows that only about 0.002 electrons are transferred from C atom to O atom at the surface. However, the calculated binding energy for this interface structure is high (to -475 meV per C atom), indicating the interfacial interaction is not likely dominated by van der Waals effect. Large interfacial distance, weak charge transfer, and high binding energy suggest that  other mechanisms such as orbital hybridization might be dominant at the interface between graphene and SrTiO$_3$. 

\begin{figure}
\includegraphics[width=2 in]{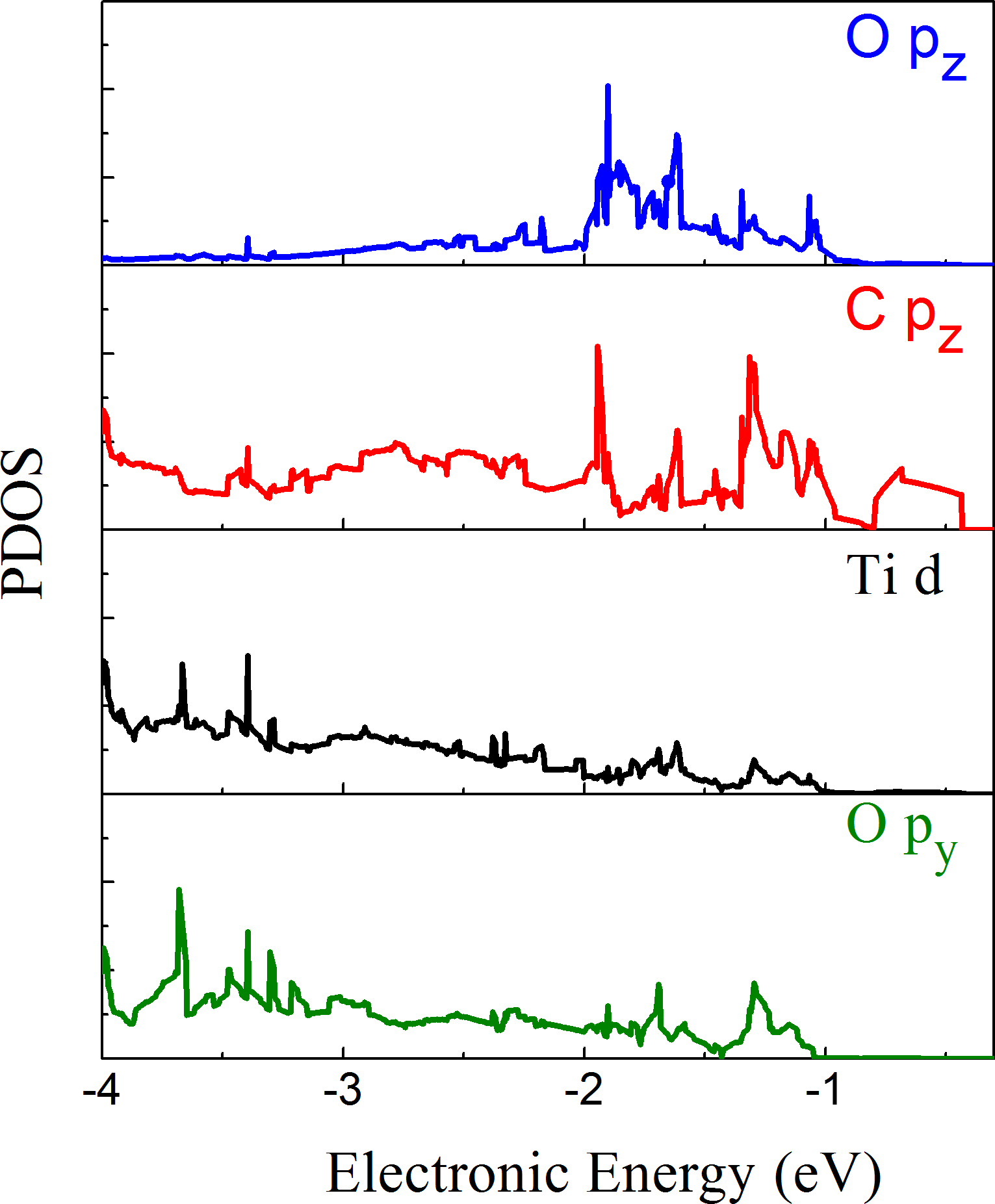}
\caption{\label{fig7} (Color online).
PDOS plots for $p_z$ orbital of C atom in graphene, and $p_y$ and $p_z$ orbital of O atom and $3d$ orbital of Ti atom.}
\end{figure}

Projected density of state (PDOS) of $p_z$ orbital of C atom in graphene, and $p_y$ and $p_z$ orbital of O atom and $3d$ orbital of Ti atom at SrTiO$_3$ surface, respectively are shown in Fig. \ref{fig7}, which clearly shows that orbital hybridization  is mainly from $p_z$ orbital of C atom in graphene and 
$p_z$ orbital of O atom at the interface of graphene and  SrTiO$_3$, especially at the energy range from -2 eV to -1 eV, while other orbital hybridization contribution such as C-$p_z$ orbital with O-$p_y$ or Ti-$3d$ orbital is weak. This orbital hybridization makes it possible so that the bands originally only occupied by O-$p_z$ electrons now can also be occupied by C-$p_z$ electrons. It is also seen that the valence bands of graphene near $\Gamma$ point are lifted upward much, compared with the band structure of pristine graphene. In addition, the orbital hybridization also pushes the conduction bands of graphene downward slightly near the $\Gamma$ point. It should be noted that this band hybridization is not likely due to the strain effect in graphene.  Most important results from these calculations are discussed further in detail in Section \ref{sec6}.

\section{Optical conductivity: Bethe-Salpeter equation (BSE) calculation details} \label{sec5}

Electron-hole interactions in the response function are inserted by solving the BSE for the  two particle correlation function  $L$,
\begin{equation}
 L = L_0 + L_0 \Xi L
 ,
 \label{BSE}
\end{equation}

Where $L_0$ is the non-interacting two particle correlation function and $\Xi$ is the BSE kernel, which is approximated as  $\Xi=-iv+iW$ ($v$ is the Coulomb potential and $W$ is the static screened potential). The optical conductivity is calculated as  $\sigma=1+i  \omega/4\pi \varepsilon_M $  where $\omega$ is the frequency,  $\varepsilon_M$ is the macroscopic dielectric function. The macroscopic dielectric function is defined in  the reciprocal space $q$ by:

\begin{equation}
 \varepsilon_{M}(\omega) = 1 -\lim_{q \to 0} v(q) L(q,\omega).
\label{diel_tens}
\end{equation}

In order to perform these many body perturbation theory BSE calculations, we first repeat the DFT- ground state calculations using Quantum Espresso code \cite{Giannozzi}. It is observed that the results from VASP and Quantum Espresso code are similar. Graphene/SrTiO$_3$  is modelled with $(\sqrt{3}\times 3)$ graphene supercell which is strained and placed on $(1 \times2)$ SrTiO$_3$ (100) surface with TiO$_2$ termination and 5 layers of thickness. About 7.61\% compressive and 6.69\% tensile strains are applied along $a$ and $b$ lattice directions of the graphene supercell, respectively, to match the lattice constants of SrTiO$_3$ substrate. In this case we employ Norm--Conserving Martins--Troullier pseudo-potentials with semicore $sp$ states for Ti and Sr atoms. The BSE calculations for absorption spectrum are performed by using BerkeleyGW package \cite{Deslippe, Onida, Strinati}.  Monkhorst--Pack (MP) mesh grid is used up to 32$\times$32$\times$1 for the RPA calculations. In order to estimate the BSE kernel, we use 8$\times$8$\times$1 MP coarse mesh grid wavefunction which is interpolated with a 16$\times$16$\times$1 finer grid in order to provide the optical conductivity. The cut off energy for the full dielectric matrix is 1.7 Ry. The total number of bands taken into account are 20 in the valence and 52 in the conduction. The absorption spectrum is calculated with the Haydock recursion method. In order to plot the excitonic e-h pair density, the two particle Hamiltonian is diagonalized with less valence and conduction bands for the finer mesh-grid wave-functions, respectively 15 and 22 bands. In this case only the resonant part is considered (Tamm-Dancoff approximation). This approximation has been demonstrated to be  accurate in the BSE calculations of absorption spectra for both the graphene monolayer \cite{Yang}and SrTiO$_3$ bulk \cite{Sponza}. 

Results from these BSE calculations are discussed in detail in Section \ref{sec6}.

\section{Results and Discussion}\label{sec6}

\begin{figure}
\includegraphics[width=\columnwidth]{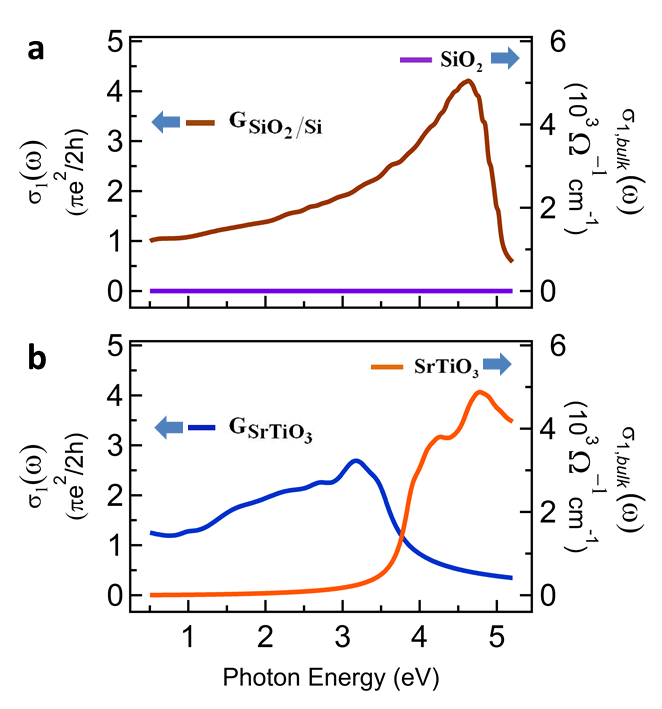}
\caption{\label{fig8} (Color online).
Real part of the optical conductivity ($\sigma_1 (\omega)$). (a) The axis in left is for sheet conductivity, $\sigma_1 (\omega)$ of G$_{SiO_2/Si}$  (in brown) while the right axis is for the bulk conductivity of SiO$_2$ (in indigo). (b) The axis in left is for sheet conductivity, $\sigma_1 (\omega)$ of G$_{SrTiO_3}$ (in blue) while the right axis is for the bulk conductivity of SrTiO$_3$ (in orange).}
\end{figure}

The real part of the optical conductivity ($\sigma_1(\omega)$) extracted from SE data for the energy range of 0.5 - 5.2 eV for G$_{SiO_{2}/Si}$ and G$_{SrTiO_3 }$ are shown in Figs. \ref{fig8}(a) and \ref{fig8}(b) respectively. The $\sigma_1(\omega)$ for   G$_{SiO_{2}/Si}$ shows characteristic graphene features observed in case of exfoliated as well as chemical vapour deposited (CVD) graphene on substrates like SiO$_2$/Si, quartz \cite{Mak11, Mak14, Kravets, Gogoi, Santoso14, Nelson, Weber} and also in free-standing graphene \cite{Chae, Nair}. From 0.5 to $\sim$1.5 eV $\sigma_1(\omega)$ is almost constant ($\pi e^{2}/2h$) which is a signature of the linear band-structure of graphene \cite{CastroNeto, Nair}. Beyond this range $\sigma_1(\omega)$ starts to gradually increase and a prominent peak is observed at 4.6 eV. This peak is attributed to optical transitions at the M point in the Brillouin zone of graphene, which is a van Hove singularity \cite{Yang, Taft, Bassani}. The peak position and the asymmetric line shape are due to e-e and e-h interactions \cite{Mak11, Mak14,  Kravets, Chae, Gogoi, Yang}.

Remarkably in the case of G$_{SrTiO_3}$, as shown in Fig. \ref{fig8}(b), in contrast we see a conspicuously different $\sigma_1(\omega)$  from that of G$_{SiO_2/Si}$ or, for that matter free-standing graphene \cite{Chae, Nair}. The most prominent change is the almost full quenching of $\sigma_1(\omega)$ after peaks at  $\sim$3.2 eV.  Another important difference is the increase of $\sigma_1(\omega)$  at lower energies ($0.5$ eV to $\sim$3.2 eV). We also observe some structures in this range whereas $\sigma_1(\omega)$ for G$_{SiO_2/Si}$ is found to be smoothly increasing gradually beyond $\sim$1.5 eV. Noting that the spectral weight for G$_{SrTiO_3}$ is not conserved as compared to G$_{SiO_2/Si}$, it may be so that the remaining spectral weight is getting transferred to states at higher energies \cite{Santoso11}.

The bulk optical conductivity ($\sigma_{1,bulk}(\omega)$) of the substrates for each case is also plotted. In Fig. \ref{fig8}(a) there is no signature of optical transitions in the substrate (for SiO$_2$) for the measured energy range. In contrast there are considerable optical transitions starting from  $\sim$3.5 eV when the substrate is SrTiO$_3$ as seen in Fig. \ref{fig8}(b).

\begin{figure*}
\includegraphics [width = 4.5 in]{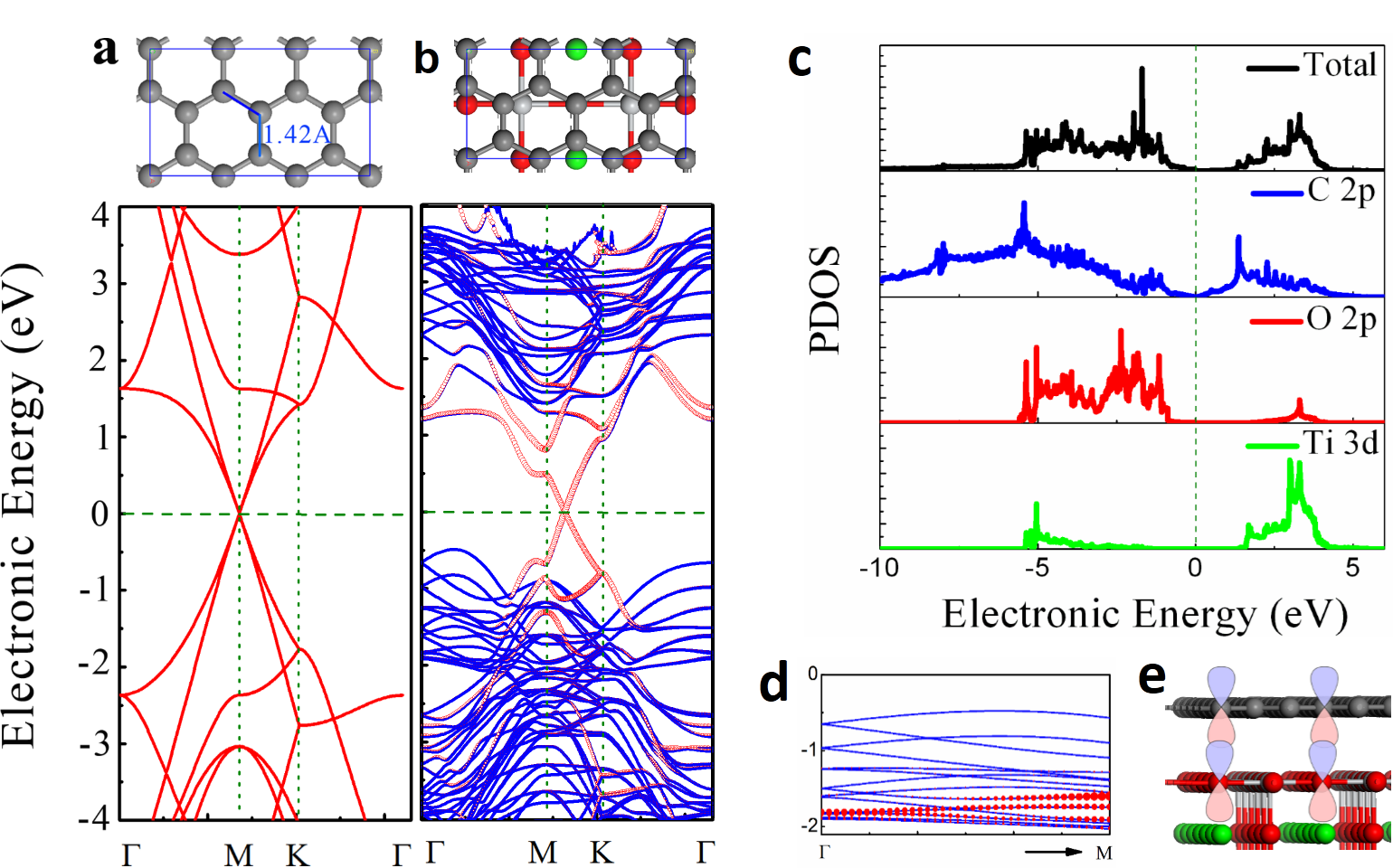}
\caption{\label{fig9} (Color online).
Ground state band-structure from DFT based calculations. (a) Band-structure for free standing  ($\sqrt 3\times 3$)   graphene supercell (top view of the supercell is in the panel above). (b) Band structure  of  graphene/SrTiO$_3$ with strained monolayer graphene on top of  2.5 unit cells (5 layers) of SrTiO$_3$ substrate. The dotted red line in the band-structure (bottom panel) denotes the contribution of C-$p_z$  bands of graphene. (c)  Projected  density of state (PDOS) plots for the various elemental orbitals.  (d)  Zoomed in band-structure of graphene/SrTiO$_3$  showing the new graphene C-p$_z$  bands not seen in that energy range for the case of freestanding graphene. The red dotted lines denote  C-$p_z$ bands. (e)  Schematic of hybridization of C-$p_z$  orbital of graphene with O-$p_z$  orbital of SrTiO$_3$ in graphene/SrTiO$_3$.}
\end{figure*}

Figures \ref{fig9}(a) and \ref{fig9}(b) show the band structures of freestanding graphene and graphene/SrTiO$_3$ respectively found using ground state DFT calculations as explained in Section \ref{sec4}. The flat bands near the M point in pristine graphene are mapped to the   $\mathit{\Gamma}$ point in the present case because of band folding. Comparing the band-structures of free standing graphene  and graphene/SrTiO$_3$,  the C-\textit{p$_z$} valence bands of the later are lifted up (Fig. \ref{fig9}(d)) significantly near the $\mathit{\Gamma}$ point, while the conduction bands are pushed down slightly. The nature of the top of the valence bands as can be seen from the PDOS plots in Fig \ref{fig9}(c) is such that there is strong overlap of the C-\textit{p$_z$} and O-\textit{p$_z$} states as explained in Section \ref{sec4}. Wave-function analysis shows strong hybrization between these  C-\textit{p$_z$}  and O-\textit{p$_z$} orbitals (schematically shown in Fig \ref{fig9}(e)).  The first structure in the conduction band as seen in the  PDOS is mainly contributed by Ti-3\textit{d} $t_{2g}$ orbitals of SrTiO$_3$ and  $\pi$* orbitals of graphene. The overall features of the PDOS of the orbitals from elements constituting the SrTiO$_3$ substrate reveal strong analogies with SrTiO$_3$  bulk  DFT calculations \cite{Sponza}.  However the PDOS of the C-\textit{p$_z$}  orbitals are modified in comparison to the pristine case.

Now, the role of e-h interactions  is revealed by the results from  BSE calculations, details of which have been explained in Section \ref{sec5}.  We compare the results of BSE calculations with those of RPA (random phase approximation) calculations, which do not take into account e-h interactions, for graphene/SrTiO$_3$. 

\begin{figure}
\includegraphics[width=\columnwidth]{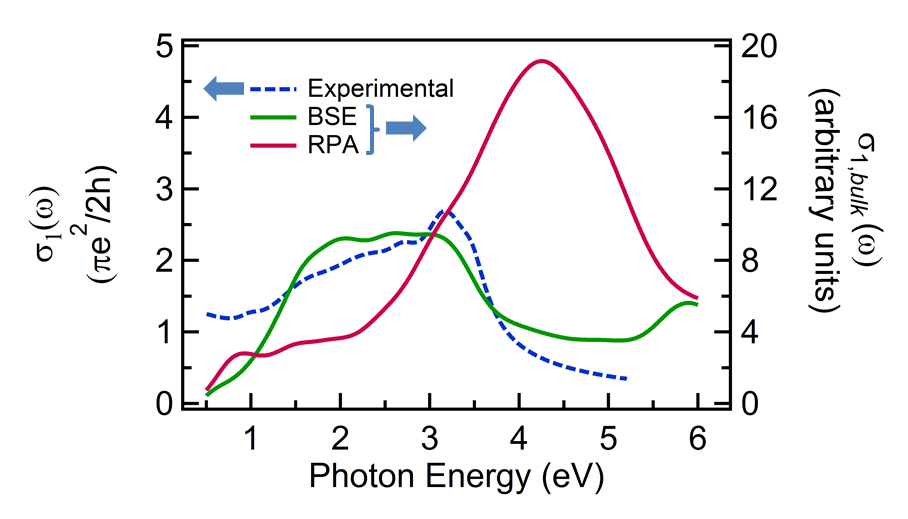}
\caption{\label{fig10} (Color online).
Calculated real part of the optical conductivity of the hybrid structure with e-h interactions using BSE (green line) and without e-h interactions using RPA (purple line). The dashed blue line represents the real part of the experimental sheet conductivity, $\sigma_1 (\omega)$ of G$_{SrTiO_3}$ }
\end{figure}

In Fig. \ref{fig10} we have plotted the real part of the optical conductivity (for graphene/SrTiO$_3$) both from RPA (red line) as well as BSE (green line). These  optical conductivities are the bulk conductivities of the interfacial system graphene/SrTiO$_3$ which have to be distinguished from the sheet conductivity of the graphene layer only. In the same plot the experimental sheet conductivity of G$_{SrTiO_3}$ i.e., the graphene layer on SrTiO$_3$ has been plotted (blue dashed line) for comparison.

We start our discussion on RPA calculations which do not take into account e-h interactions. The RPA optical conductivity results show a prominent peak at 4.3 eV which originates from the van Hove singularity of hybridized $\pi-\pi^*$ graphene orbitals with O-\textit{2p} and Ti-\textit{3d} from SrTiO$_3$.  At lower energy the RPA optical conductivity shows a typical graphene behavior originating from the $\pi-\pi^*$ graphene bands. This RPA result clearly does not follow experimental result. The zero in optical conductivity results at low energy ($\sim$0.5 eV) is an artefact of the finite k-mesh grid which has been used.

\begin{figure*}
\includegraphics[width= 4.5 in]{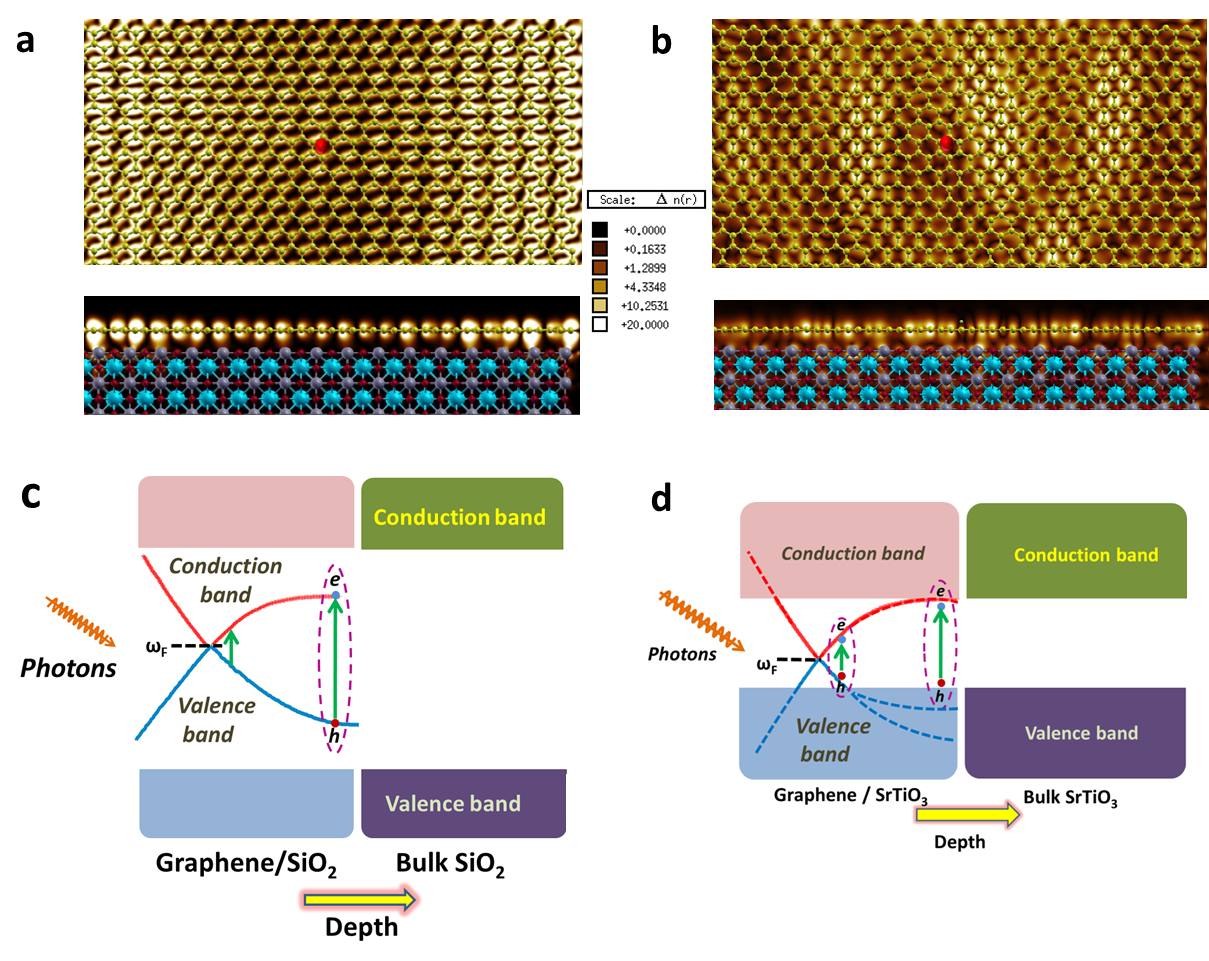}
\caption{\label{fig11} (Color online).
Electron – hole (e-h) density plots and schematic e-h interaction mechanism. (a) Top-view (upper panel) and side- view (lower panel)  of electron density for e-h excitation  at energy 1.5 eV. Here gold, dark red, light blue, light violet balls represent carbon, oxygen, strontium and titanium atoms respectively. The red dot represents the position of the hole on the graphene layer.  (b) Top-view (upper panel) and side- view (lower panel) of electron density for e-h excitation at energy 3.1 eV. The color scheme is the same as in a.  (c) Schematic band diagrams for the top layer (shown as hybrid graphene/SiO$_2$) and the bulk substrate underneath for graphene/SiO$_2$/Si system. The resonant excitonic effects are present in purely graphene bands (till our energy range of concern). (d) Schematic band diagrams for the top layer (shown as hybrid graphene/SrTiO$_3$) and the bulk substrate underneath for graphene/SrTiO$_3$ system. In contrast to the previous case the resonant excitonic effects in graphene are strongly affected by SrTiO$_3$  substrate, whose band-gap is far smaller than SiO$_2$. The excitonic wavefunctions involve both graphene and substrate orbitals.  }
\end{figure*}

The striking result comes from BSE calculations which take into account e-h interactions.  As one can immediately see the RPA picture is severely modified with the introduction of  e-h interactions within the BSE framework.  The BSE result shows considerable red-shift of spectral weight to the photon energy range between $\sim$1 and $\sim$3 eV. Remarkably optical conductivity decreases conspicuously starting from $\sim$3.2 eV which agrees well with the experimental sheet conductivity of G$_{SrTiO_3}$. The close agreement between the experimental and theoretical results emphasizes the fact that although our calculations are based on the total system graphene/SrTiO$_3$ still the resultant bulk conductivity is dominated more by graphene contributions. Our detailed analysis show that the transitions which occur above $\sim$3.2 eV in BSE calculations mostly come from SrTiO$_3$ or in other words from the \lq bulk\rq\,  SrTiO$_3$ substrate. This could be observed prominently in the calculated optical conductivity beyond $\sim$3.5 eV as shown in Fig. \ref{fig10}.  This scenario has been confirmed, by looking at the nature of the  excitonic wave functions. In Fig. \ref{fig11}(a) and \ref{fig11}(b), we have plotted the excitonic e-h denisty for excitations at 1.5 eV and 3.1 eV respectively for each atoms ( O, Ti of SrTiO$_3$  and C of graphene ). The hole is placed at the center of graphene layer (red dot).

In Fig. \ref{fig11}(a) for 1.5 eV excitations, the electron density is mainly associated to the $\pi$* graphene orbitals. Surprisingly, even for these low energy excitations (below the band gap of bulk SrTiO$_3$), the Ti-\textit{3d}  orbitals in the first layer of SrTiO$_3$ interface are also involved.  Below $\sim$1.5 eV the increased optical conductivity from the universal value (as observed in the case of freestanding graphene as well as graphene on SiO$_2$/Si etc.) could be due to the different role that e-e interactions play in graphene/SrTiO$_3$ but not taken into account in the DFT calculation. The excitonic wave function density for the electrons associated with the intensive excitation at 3.1 eV, shown in Fig. \ref{fig11}(b), now consists of  the $\pi$* graphene orbitals with \emph{bulk} Ti-\textit{3d}  states of the SrTiO$_3$. At higher energies, the distinction between intrinsic graphene and SrTiO$_3$ features becomes subtle with the excitonic wave functions however mostly involving SrTiO$_3$ orbitals.

In Fig. \ref{fig11}(c) and \ref{fig11}(d) we schematically summarize the mechanisms involved for both the systems graphene/SiO$_2$/Si and graphene/SrTiO$_3$. As monolayer graphene is a two dimensional (2D) material, the whole layer itself could have interfacial properties together with one or a few layers of the substrate. The electronic structure of the underlying substrate plays the most crucial role in controlling the nature of this new hybrid system.  In the case of graphene/SiO$_2$/Si we are in the \textit{intrinsic} graphene  regime as the substrate bands are far away from the substrate in contact, SiO$_2$ which  has a band gap of $\sim$9 eV, thus the graphene-SiO$_2$/Si interactions are weak. Our results from SE confirm that G$_{SiO_2/Si}$ shows intrinsic graphene-\textit{like} character. On the other hand in Fig. \ref{fig11}(d) we show the case of graphene/SrTiO$_3$.  Here the smaller bulk band gap of SrTiO$_3$ modifies the scenario strongly with two main effects. Firstly the ground state wave-functions of graphene are hybridized with those of the SrTiO$_3$ substrate. Secondly, the optical conductivity is strongly affected by resonant excitonic effects. The e-h interactions involve electron and hole states from both graphene and SrTiO$_3$. For the graphene layer this coupling through resonant excitonic effects is strongest when the excitation energy is near the bulk band gap energy of SrTiO$_3$.  The experimental result for the optical conductivity of the graphene layer on SrTiO$_3$ (G$_{SrTiO_3}$) with drastic reduction beyond the UV range ( $> \sim$3.2 eV) supports these descriptions. It is then important in future to study how the band gap of substrates with other intermediate band gaps affect the optical conductivity  of graphene.

\section{conclusion}
Overall this work elucidates novel and important physics of Ti-3\textit{d} orbital in SrTiO$_3$  interacting  with graphene, modifying its optical conductivity. Our results and methodology open a new path of manipulating many-body interactions in graphene via \textit{3d} orbitals and can be extended to others systems such as Mott-insulator.

\section{Acknowledgements}

We acknowledge Shaffique Adam and Vitor M. Pereira for the discussions and their valuable comments. This work is supported by Singapore National Research Foundation under its Competitive Research Funding (NRF-CRP 8-2011-06, R-144-000-295-281 and NRF2008NRF-CRP002024), MOE-AcRF Tier-2 (MOE2010-T2-2-121), NUS-YIA, FRC, and NUS Core Support C-380-003-003-001. We acknowledge the CSE-NUS computing center and Graphene Research Center for providing facilities for our numerical calculations. P.K.G., P.E.T. and M.Y. contribute equally to this work.
%

%

%
%
\end{document}